\documentclass[useAMS,usenatbib]{mn2e}
\usepackage[dvips]{graphicx}
\usepackage{amsmath}

\title[Colour gradients in normal and compact early-type galaxies at 1$<$z$<$2]
{Colour gradients in normal and compact early-type galaxies at 1$<$z$<$2}
\author[A. Gargiulo, P. Saracco,
  M. Longhetti]{A. Gargiulo$^{1}$\thanks{E-mail:
    adriana.gargiulo@brera.inaf.it}, P. Saracco$^{1}$,
  M. Longhetti$^{1}$\\
\newline
$^{1}$INAF - Osservatorio Astronomico di Brera,via Brera 28, 20121, Milano, Italy}

\begin{document}

\date{Accepted 1988 December 15. Received 1988 December 14; in original form 1988 October 11}

\pagerange{\pageref{firstpage}--\pageref{lastpage}} \pubyear{2002}

\maketitle

\label{firstpage}

\begin{abstract}
We have derived colour gradients 
for a sample of 20 early-type galaxies (ETGs) at $1<z_{spec}<2$ selected from
the GOODS-South field.
The sample includes both normal ETGs (13) having effective radii 
comparable to
the mean radius of local ones and compact ETGs (7) having effective radii
from two to six times smaller.
Colour gradients have been derived in the F606W-F850LP bands (~UV-U 
rest-frame)
taking advantage of the ultradeep HST-ACS observations covering this field
and providing a spatial resolution of about 0.8 kpc at the redshift of the
galaxies.
Despite of the narrow wavelength baseline covered (1000 \AA), sampling
approximatively the emission dominated by the same stellar population, we
detect significant radial colour variations in 50 per cent of our sample.
In particular, we find five ETGs with positive colour gradients (cores bluer
than the external regions), and five galaxies with negative colour gradients
(cores redder than the external regions), as commonly observed in the local
Universe. These results show that at $1<z<2$, when the Universe was only 
3-4
Gyr old, ETGs constituted a composite population of galaxies whose
different assembly histories have generated different stellar distributions
with the bluest stellar population either in the center or in the outskirts as well
as throughout the whole galaxy. Moreover, we find that
compact galaxies seem to preferentially show a blue cores while moving
towards normal galaxies, central stellar populations become progressively
redder. Nonetheless, the narrow baseline covered together with the
low statistics still prevent us to be conclusive about a possible physical
connection between colour gradients and the degree of compactness of high-$z$ ETGs.

\end{abstract}

\begin{keywords}
galaxies: elliptical and lenticular, CD galaxies:formation, galaxies: evolution, galaxies: high-$z$, galaxies: stellar content
\end{keywords}

\section{Introduction}

A powerful tool to investigate how stellar mass is accreted onto ETGs
from their formation until now, is to resolve the spatial distributions of
their stellar populations and observe how they changed with
time. Indeed, colour gradients are the most direct measure that can
provide us with information on the stellar distribution within a
galaxy, but until now, instrumental limits have restricted their study
only to the local and intermediate Universe.  In the last years, the
advent of the HST and its capability into resolve distant galaxies are
opening new possibilities into estimating colour gradients even for
high-$z$ galaxies.  Actually, the lack of multiband imaging has
prevented an effective analysis of colour gradients of high-$z$ ($z>1$) ETGs in
all but two cases.  \citet{mcgrath08} derived the colour-maps for 2
ETGs at $z\sim$1.5 in the F814W-F160W HST bands, finding flat
gradients out to 2R$_e$.  Some years before, \citet{moth02} studied the
rest-frame UV218-U300 colour gradient for two samples of
33 galaxies at 0.5$<z<$1.2 and 50 galaxies at 2.0$<z<$3.5 in the 
Hubble Deep Field North. In fact, this analysis even representing the 
widest study of colour gradients of high-$z$ galaxies until then,
is limited by the presence of photometric redshift for half of the high-$z$ 
sample, and suffers of the large pixel scale of HST-NIC3 camera which provides a
spatial resolution of 1.6 kpc at $z\sim$ 1.5 insufficient to
clearly assess the morphology of the galaxies and properly investigate 
colour gradients at R $<$ R$_e$. Nonetheless, in their composite samples they found
an inversion in the colour trend from negative values (stars redder in the center)
at lower redshifts to a positive values (stars bluer in the center) at higher redshifts.
Actually, positive colour
gradients were also detected in a sample of pure spheroidals at 
intermediate redshifts \citep{menanteau01,menanteau05}, suggesting an inversion of colour
gradients from negative values generally observed in the local
Universe to positive values at higher redshift. In fact, the first
studies on high-$z$ ETGs are showing that local and early ETGs
populations differ not only for the radial colour variations, but that
the whole picture is more complex than the homogeneous one
hypothesized from studies of local ETGs.

Recent observations \citep{saracco09,mancini10} have shown that at
high-$z$ (1$<z<$2), when the Universe was only 3-4 Gyr old, ETGs
(hereafter we say compact) with effective radius 2-3 times smaller
than that of a typical present-day ETG coexist with a majority of
early-type galaxies whose dimensions are similar to those of local
ETGs (we say normal). The only attempt to measure the velocity
dispersion of a high-$z$ ($z$=2.186) compact ETG \citep{vandokkum09}
has resulted in a $\sigma$ value of 510$^{+165}_{-95}$ suggesting in
these galaxies a denser distribution of stellar mass than in normal
ETGs. At the same time, the few estimates of velocity dispersions
recovered for normal high-$z$ ETGs confirmed their similarity with local
populations even for stellar mass density
\citep{cenarro09,cappellari09,onodera10}.  These findings point out
that the way in which the stellar mass was accreted on ETGs in the
first 3-4 Gyr, is not univocal for the whole population, but on the
contrary, the differences both in dimension and in structure of
compact and normal high-$z$ ETGs suggest that they underwent different
scenarios of formation and/or early evolution \citep{saracco10}.

Moreover, the compact ETGs, at first sight, seemed to be very rare in
the local Universe \citep{shen03,trujillo09} and this evidence
corroborates the idea that compact and normal galaxies should have
different mass assembly histories even in the following 9-10
Gyr. Indeed, differently from normal galaxies, compact ETGs should
undergo a size evolution to reconcile them with the local dimension of
a typical local ETGs. Many scenarios were proposed to match the
size/density of high-$z$ and local ETGs e.g. non dissipative ``dry''
merger, adiabatic expansion, age/colour gradient
\citep{boylan06,naab09,fan08,damjanov09,labarbera09}.  Among these,
wide credit is given to the merger hypothesis \citep{naab09,hopkins10}
even if no result is still conclusive \citep{nipoti09}.  Additionally,
in the last year, a new hypothesis is rising beside these
scenarios. Indeed, compact galaxies with masses and radii comparable
with those found at high-$z$ are now detected both in local and
intermediate galaxy clusters
\citep{valentinuzzi10,valentinuzzi10b} as well as in the
field \citep{stockton10} and the number density of compact galaxies in
local clusters turns out to be consistent with the one measured at
high-$z$ \citep{saracco10}. This unexpected results have shaken the
previous belief whereby compact galaxies were physical systems
confined to the early Universe and that they must subsequently undergo
an apparent or real size evolution. Indeed, the recent observations
have shown that even the local population of ETGs is composed, as in
the early Universe, by normal galaxies which obey the well studied
scaling relation and by a fraction of compact galaxies that, with
their small radius, and enhanced stellar mass density fall out of any
correlations.

From these observations, it is clear that to gain insight in the
complex picture of ETGs formation and evolution, it is necessary to
individually define the mass assembly paths followed by both normal
and compact ETGs. To address this topic it turns out to be fruitful to
study the evolution of the stellar content of both compact and normal
galaxies from the early Universe until the present day, and investigate if any
difference occurs. To this aim, in this paper we present the study
of colour gradients for a composite sample of both compact and normal
high-$z$ ETGs. This work ascertains the feasibility of this analysis even
for high-$z$ ETGs, and provides us the first direct information on the
distribution of the stellar content in the first stages of their
life. 

In Section 2, we present our sample, and in Section 3 we
describe the methods used to derive colour gradients for high-$z$
ETGs. In Section 4 we show our results and in Section 5 we present our
conclusions.

Throughout this paper we adopt a standard cosmology with
H$_{0}$\,=\,70\,km\,s$^{-1}$\,Mpc$^{-1}$, $\Omega_{m}$\,=\,0.3 and
$\Omega_{\Lambda}$\,=\,0.7. All the magnitudes are in AB system. 

\section {The sample}

An exhaustive study of the colour gradients in high-$z$ ETGs until now has
been severely limited by many factors such as the lack of spectroscopic
redshifts, of a morphological classification, and of images with high
resolution and signal to noise.  Indeed, in the redshift range we are
interested in (1$<z<$2), the redshift estimates are strictly
compromised by the lack of prominent features in the optical spectra,
and as pointed out by \citet{mancini10}, a reduced S/N can affect the
estimate of the effective radius due to the lack of possible
peripheral structures with low surface brightness such as wings or
halos. In fact, the minimum spatial resolution and the signal-to-noise
ratio necessary to properly model the light profiles from at least the
first central kpc to regions beyond the effective radius, is achievable
only with deep space-based imaging. Till now, these constrains have strongly
limited the collection of a representative sample able to address this
topic from the already meagre observations available for high-$z$ ETGs.

We collected a sample of 20 ETGs at 0.9$<z_{spec}<$1.92 best suited
for our purpose (see below), extracted from the complete sample of 34
ETGs selected from the southern field of the Great Observatories
Origins Deep Survey (GOODS-South v2; Giavalisco et
al. \citeyear{giavalisco04}) by \citet{saracco10}.  The survey
provides each galaxy both with deep HST-ACS imaging in four bands
(F435W, F606W, F775W and F850LP) and with spectroscopic redshift
(Vanzella et al.  \citeyear{vanzella08} and references therein). In
addition, surface brightness parameters in the F850LP band, age and
stellar mass M$_{\star}$ derived in an homogeneous way for the whole
sample are available.  For more details on the selection criteria,
morphological classification, stellar mass and age estimations of the
original complete sample, see \citet{saracco10} and references
therein.

Among the four bands at our disposal, we opted to measure the
F606W-F850LP colour variations corresponding approximatively to UV-U
colour in the rest-frame in order to obtain a compromise between the
wavelength baseline covered and the S/N achieved. Actually, the F453W
images have a S/N $\sim$ 2 times lower than that of F606W images, and
this would have prevented us from a reliable analysis of the surface
brightness parameters for most of the sample.  From the original
sample of 34 ETGs, we removed 3 galaxies showing X-ray emission and 11
galaxies showing anomalies in the analysis of the F606W-band
images. In particular, these comprise 6 galaxies whose fitting light
profile does not reproduce the observed ones, that is showing
significant residuals in the residual maps, and 5 galaxies whose S/N
was too low to derive their surface brightness profile out to
R$_e$. For the remaining 20 galaxies, the FWHM of $\sim$ 0.1" of both
F606W- and F850LP-band HST-ACS images, corresponding to a spatial
resolution element of $\sim$0.8 kpc at z=1.5, and the exposure times
ranging from $\sim$ 20 ks to 50 ks, and from $\sim$ 35 ks to 100 ks
for the F606W and F850LP filters, respectively, allow us to accurately
define the light profile from the innermost regions to well beyond
R$_{e}$.

\begin{figure}
	\includegraphics[angle=0,width=8.5cm]{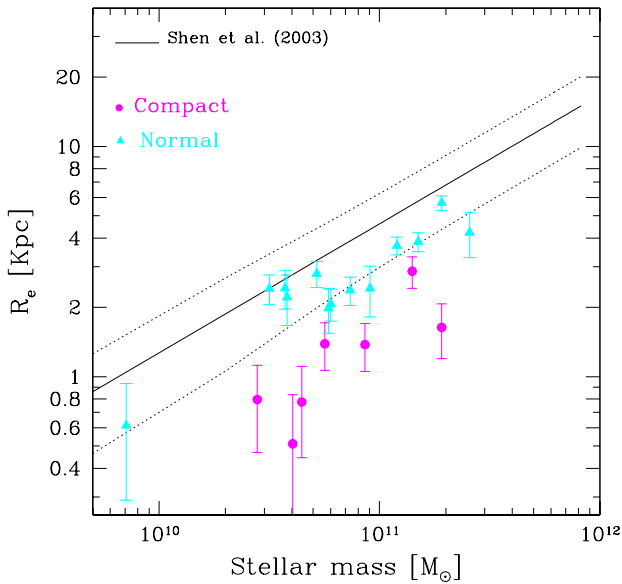} \\
	\caption{Size-mass relation for local ETGs (solid line, Shen
          et. al 2003) and for our sample (solid symbols). The dashed
          lines are the scatter lines at 1$\sigma$. We shifted the
          Shen et al.'s relation by a factor $\sim$\,1.2 towards lower masses
          to take into account the systematic shift observed in the
          mass estimations using our models or those adopted by Shen
          et al. (2003). The circles are compact galaxies,
          i.e. galaxies having effective radius more than one sigma
          smaller than those predicted by local relation for that
          mass. On the contrary, galaxies having effective radius
          comparable at 1$\sigma$ with those aspected by Shen et al.'s
          relations, are classified as normal (triangle symbols).}
	\label{sm}
\end{figure}

Following \citet{saracco10} we define ``compact" ETGs those galaxies
that, in the F850LP band, lie more than one sigma below the local
size-mass (SM) relation by \citet{shen03}, i.e. the relation between
the effective radii R$_{e}$ and the stellar masses.  To visually
observe how our galaxies are distributed respect the SM relation, and
hence their degree of compactness, in Fig. \ref{sm}
we report the SM relation in the F850LP band for our sample (solid
symbols) and for local galaxies (solid line, Shen et al.  2003). The
dashed lines represent the scatter at 1$\sigma$. It turns out that 7
out of 20 galaxies (solid points) of our sample are ``compact". In
order to have a quantitative estimates of the degree of compactness,
we used the ratio R$_{e}$/(R$_{e,z=0}$-1\,$\sigma$) where R$_{e}$ is
the effective radius of the galaxy and R$_{e,z=0}$ is the effective
radius that a galaxy of equal mass would have at $z$\,=\,0 as derived
by the local SM relation. In table \ref{sp} we report the compactness
values for our galaxies. It is to note that, according to this
definition, compact galaxies have R$_{e}$/(R$_{e,z=0}-1\,\sigma)<$0.9
and lower value of degree of compactness corresponds to galaxies more
compact.

Thus, after the selection, our sample is formed
by 20 ETGs at 0.9$<z_{spec}<$1.92, 7 compact and 13 normal, 
with stellar masses ranging from $\sim$
10$^{10}$\,M$_{\odot}$ to $\sim$ 3$\times$10$^{11}$\,M$_{\odot}$, and age
varying from $\sim$ 1 Gyr to $\sim$ 3.5 Gyr.

\section{Estimate of colour gradient of ETGs}

Following previous works (e.g. Peletier et al. \citeyear{peletier90};
Saglia et al. \citeyear{saglia00}), to obtain a quantitative estimate
of colour variation $\nabla$(UV-U)$_{restframe}$ along a fixed galaxy
radius we measured the logarithmic slope of the galaxy colour profile
$\mu_{UV}$(R)-$\mu_{U}$(R):
\begin{equation}
\nabla_{UV-U} = \frac{\Delta (\mu_{UV}(R)-\mu_{U}(R))}{\Delta \log R}
\label{sb}
\end{equation}
where $\mu_{UV}$(R) and $\mu_{U}$(R) are the surface brightness
profiles of the galaxy in the UV and U, respectively.

\subsection{Estimates of light profiles of high-$z$ ETGs}

We fitted the light profiles of our galaxies in the F606W and F850LP bands
with a Sersic profile (\citeyear{sersic68}):
\begin{equation}
\mu(R) = \mu_{e} + \frac{2.5b_{n}}{\ln(10)}[(r/r_{e})^{1/n}-1]
\label{mu}
\end{equation}
where the effective radius r$_{e}$ [arcsec], the surface brightness at
r$_{e}$, $\mu_{e}$, and the Sersic index $n$ are the three structural
parameters that shape the profile. To perform the fit we took
advantage of the software $\small{GALFIT}$ \citep[v2,][]{peng02}, a
semiautomatic tool that starting from initial parameters provided by
the user, accurately models galaxy profile by means of a
two-dimensional fit. In the fitting procedure, the software convolves
the galaxy model with the point-spread function (PSF) to take into
account the blurring of the profile due to the diffraction and
scattering of light as it passes through the telescope and instrument
optics. The impact of the PSF on the light profile regards
mainly the internal region of the profile \citep[e.g. see][]{peng02}
where it is extremely sensitive to the Sersic index $n$ and if not
taken into account can alter the shape, and consequently the slope of
the colour profile. To properly remove this effect, for each
galaxy we fitted the surface brightness parameters adopting different
PSFs. We modelled the PSFs by sampling the light profiles of some
unsaturated stars as near as possible to the galaxy, as well as by
averaging these profiles and retaining as the ``true" PSF model, the one
which returns the best residual map. The output of the software
returns to the user the total magnitude M$_{tot}$, the Sersic index
$n$, the semimajor axis $a$, and the axial ratio $b/a$, which are
related to the light profile parameters through:
\begin{equation}
r_{e} = a\sqrt{b/a}, 
\label{re}
\end{equation}
and
\begin{equation}
\mu_{e} = <\mu>_{e} + 2.5 \log (n e^{b}\Gamma(2n)/b^{2n})
\end{equation}
where $<\mu>_{e}$ = M$_{tot}$ + 2.5 $\log$ (2\,$\pi$\,r$_{e}^{2}$), b
= 2$n$ - 1/3 + 0.009876/$n$ as found by Prugniel\,$\&$\,Simien (1997),
and $\Gamma$(2$n$) is the complete Gamma function (Ciotti 1991).  To
derive the colour variation along a fixed galaxy radius, it is
necessary that the light-profiles in the two bands are evaluated on
identical ellipses both for shape and for orientation in order to
prevent any artificial gradient.  To obtain this, we run GALFIT on the
F606W images keeping fixed both the position angle and the axial ratio
b/a at the values estimated on the F850LP images. We chose the F850LP
band as reference being the one providing the highest S/N ratio.  In
some cases (3 galaxies both in F850LP- and F606W-band images and 5
only in F606W-band images) the fitting procedure is not able to
converge. For these galaxies, we re-performed the fit setting the
Sersic index to a fixed value. In particular, we run the algorithm
many times with different value of fixed $n$, and kept as final
solution the one for which the fit converges and the residual map are
lacking of any remaining features and structures. In table \ref{sp} we
report the total magnitude F606W$_{tot}$ and F850LP$_{tot}$ for the
two bands, respectively, the effective radius, and the Sersic index
$n$ with their uncertainties as estimated by $\small{GALFIT}$ for the
two bands. It is to note that $\small{GALFIT}$ errors are
known to be severely underestimated \citep[see][]{haussler07}.  The
galaxies fitted with a fixed value of $n$ can be recognized by their
Sersic index errors setted at 0.0.
\begin{table*}
\caption{Our sample of galaxies. $Column$ $1$: id number $Column$ $2$:
  spectroscopic redshift; $Column$ $3$,$4$: total magnitude from
  GALFIT in F606W ad F850LP band, respectively; $Column$ $5$,$6$:
  effective radius in kpc; $Column$ $7$,$8$: Sersic index; $Column$
  $9$: compactness; $10$: colour gradient.  Errors on magnitude and structural parameters
  are from GALFIT, while errors on colour gradients are estimated by
  simulations (see text). At the median redshift z = 1.5 1\,arcsec
  corresponds to $\sim$ 8.6\,kpc.} \footnotesize
\begin{tabular}{ccccccccccc}
\hline
Object & z & F606W$_{tot}$ & F850LP$_{tot}$ & R$_{e,606}$ & R$_{e,850}$ & n$_{606}$ & n$_{850}$ & Compactness                    & $\nabla_{UV-U}$ \\
       &   & mag           & mag            & kpc         & kpc         &           &           & R$_{e}$/(R$_{e,z=0}$-1$\sigma$) &  mag/dex\\
\hline
12965 & 1.02 & 25.65$\pm$0.04 & 23.26$\pm$0.02 & 0.54$\pm$0.03 & 0.61$\pm$0.02 & 3.3$\pm$0.5   & 3.8 $\pm$0.2  & 1.21  &  0.0 $\pm$0.2\\
11888 & 1.04 & 23.83$\pm$0.06 & 21.51$\pm$0.01 & 3.6 $\pm$0.4  & 2.37$\pm$0.05 & 5.4$\pm$0.3   & 4.78$\pm$0.06 & 1.07  & -0.19$\pm$0.08\\
11539 & 1.10 & 23.40$\pm$0.05 & 20.7 $\pm$0.2  & 4.6 $\pm$0.4  & 4.2 $\pm$0.4  & 4.5$\pm$0.2   & 4.0$\pm$0.1   & 0.91  &  0.04$\pm$0.09\\
 9066 & 1.19 & 25.50$\pm$0.06 & 23.0 $\pm$0.4  & 2.19$\pm$0.15 & 2.43$\pm$0.16 & 2.5$\pm$0.0   & 3.5$\pm$0.2   & 1.72  & -0.3 $\pm$0.2\\
12789 & 1.22 & 25.93$\pm$0.05 & 23.09$\pm$0.05 & 0.91$\pm$0.08 & 2.2 $\pm$0.2  & 3.5$\pm$0.0   & 5.0$\pm$0.3   & 1.54  &  0.7 $\pm$0.4\\
12000 & 1.22 & 25.03$\pm$0.06 & 22.60$\pm$0.01 & 2.7 $\pm$0.2  & 2.40$\pm$0.07 & 5.0$\pm$0.0   & 5.0$\pm$0.0   & 1.90  & -0.08$\pm$0.12\\
9369  & 1.30 & 25.83$\pm$0.08 & 22.94$\pm$0.04 & 0.94$\pm$0.12 & 1.98$\pm$0.15 & 2.7$\pm$0.5   & 4.7$\pm$0.2   & 1.03  &  0.5 $\pm$0.4\\
11804 & 1.91 & 25.41$\pm$0.07 & 23.51$\pm$0.03 & 3.2 $\pm$0.2  & 2.86$\pm$0.16 & 2.5$\pm$0.0   & 4.5$\pm$0.0   & 0.87  & -0.8 $\pm$0.3\\
2     & 0.96 & 22.64$\pm$0.04 & 20.23$\pm$0.01 & 5.4 $\pm$0.3  & 5.69$\pm$0.13 & 5.3$\pm$0.1   & 5.71$\pm$0.06 & 1.45  & -0.02$\pm$ 0.09\\
13    & 0.98 & 22.72$\pm$0.02 & 20.37$\pm$0.01 & 4.6 $\pm$0.2  & 3.71$\pm$0.04 & 5.5$\pm$0.1   & 5.46$\pm$0.03 & 1.24  & -0.16$\pm$0.06\\
20    & 1.02 & 23.66$\pm$0.01 & 20.95$\pm$0.01 & 2.73$\pm$0.05 & 3.85$\pm$0.09 & 4.0$\pm$0.0   & 5.38$\pm$0.06 & 1.13  &  0.0 $\pm$0.08 \\
23    & 1.04 & 24.17$\pm$0.06 & 21.37$\pm$0.02 & 3.2 $\pm$0.3  & 2.80$\pm$0.08 & 3.4$\pm$0.2   & 3.90$\pm$0.07 & 1.59  & -0.30$\pm$0.13\\
3     & 1.04 & 23.71$\pm$0.03 & 21.52$\pm$0.01 & 1.18$\pm$0.08 & 1.39$\pm$0.02 & 7.7$\pm$0.4   & 4.87$\pm$0.06 & 0.74  &  0.43$\pm$0.04\\
11    & 1.09 & 24.86$\pm$0.04 & 22.00$\pm$0.02 & 2.15$\pm$0.13 & 2.4 $\pm$0.2  & 5.0$\pm$0.0   & 5.00$\pm$0.02 & 0.96  &  0.09$\pm$0.10\\
12    & 1.12 & 24.08$\pm$0.05 & 21.62$\pm$0.01 & 1.8 $\pm$0.2  & 1.37$\pm$0.03 & 6.5$\pm$0.4   & 4.94$\pm$0.07 & 0.56  &  0.04$\pm$0.05\\
8     & 1.12 & 24.83$\pm$0.03 & 22.02$\pm$0.01 & 1.59$\pm$0.09 & 2.07$\pm$0.04 & 5.0$\pm$0.0   & 5.00$\pm$0.00 & 1.06  &  0.21$\pm$ 0.09\\
2148  & 1.60 & 25.7 $\pm$0.3  & 23.29$\pm$0.04 & 4.6 $\pm$6.8  & 1.64$\pm$0.15 & 5.5$\pm$25.73 & 4.8$\pm$0.3   & 0.42 &  0.3 $\pm$0.4\\
2111  & 1.61 & 26.32$\pm$0.09 & 23.87$\pm$0.03 & 0.78$\pm$0.11 & 0.78$\pm$0.04 & 2.8$\pm$0.7   & 3.5$\pm$0.3   & 0.49 & -0.3 $\pm$0.4\\
2355  & 1.61 & 26.17$\pm$0.10 & 24.21$\pm$0.02 & 0.92$\pm$0.17 & 0.79$\pm$0.03 & 3.5$\pm$0.9   & 2.1$\pm$0.2   & 0.68 &  0.5$\pm$0.4\\
472   & 1.92 & 26.37$\pm$0.04 & 24.38$\pm$0.02 & 0.40$\pm$0.03 & 0.51$\pm$0.03 & 2.3$\pm$0.6   & 2.4$\pm$0.3   & 0.34 &  0.3$\pm$0.5\\
\hline
 \label{sp}
 \end{tabular}
 \normalsize
\end{table*}

\subsection{Testing the reliability of structural parameters}

\begin{figure}
	\includegraphics[angle=-90,width=8cm]{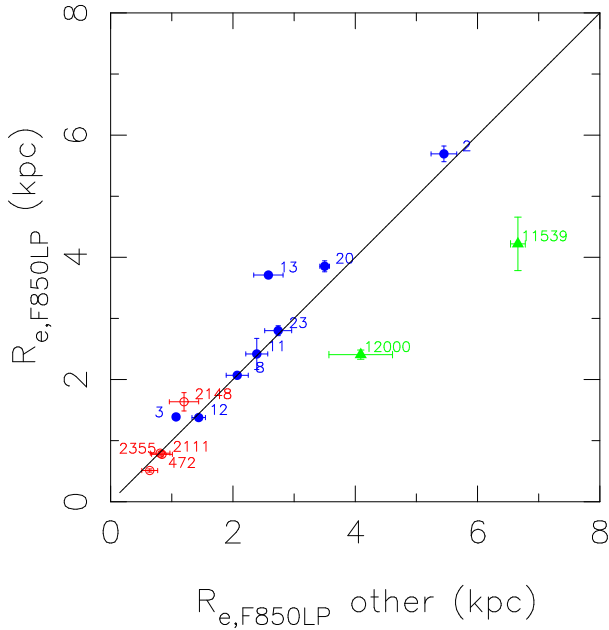} \\
	\caption{The comaprison between our estimates of R$_{e}$ and
          those of other groups in the F850LP band. Blue filled points
          are galaxies from Rettura et al., red open points are those
          from Cimatti et al., while green triangles are from
          di Serego Alighieri et al..}
	\label{comparisonre}
\end{figure}

Having derived the structural parameters, we checked how our estimates
are related to those of other studies.  To this aim, we used the 14
galaxies of our sample in common with other groups. In particular, 8
galaxies are from Rettura et al. (\citeyear{rettura06}) while 4
galaxies are from \citet{cimatti08} and 2 from \citet{diserego05}. In
Fig \ref{comparisonre} we show the comparison between our estimates of
the effective radii R$_{e}$ [kpc] measured in the F850LP band and
those by other groups.  The agreement is good (within the errors) for
$\sim$ 60$\%$ of the sample and, above all, no trend is observed in
the whole range of radius, magnitude, and redshift.

We then assured the reliability of our estimates through two different
checks: i) performing a set of simulations, and ii) comparing the
fitted and the observed light profiles.

\begin{figure*}
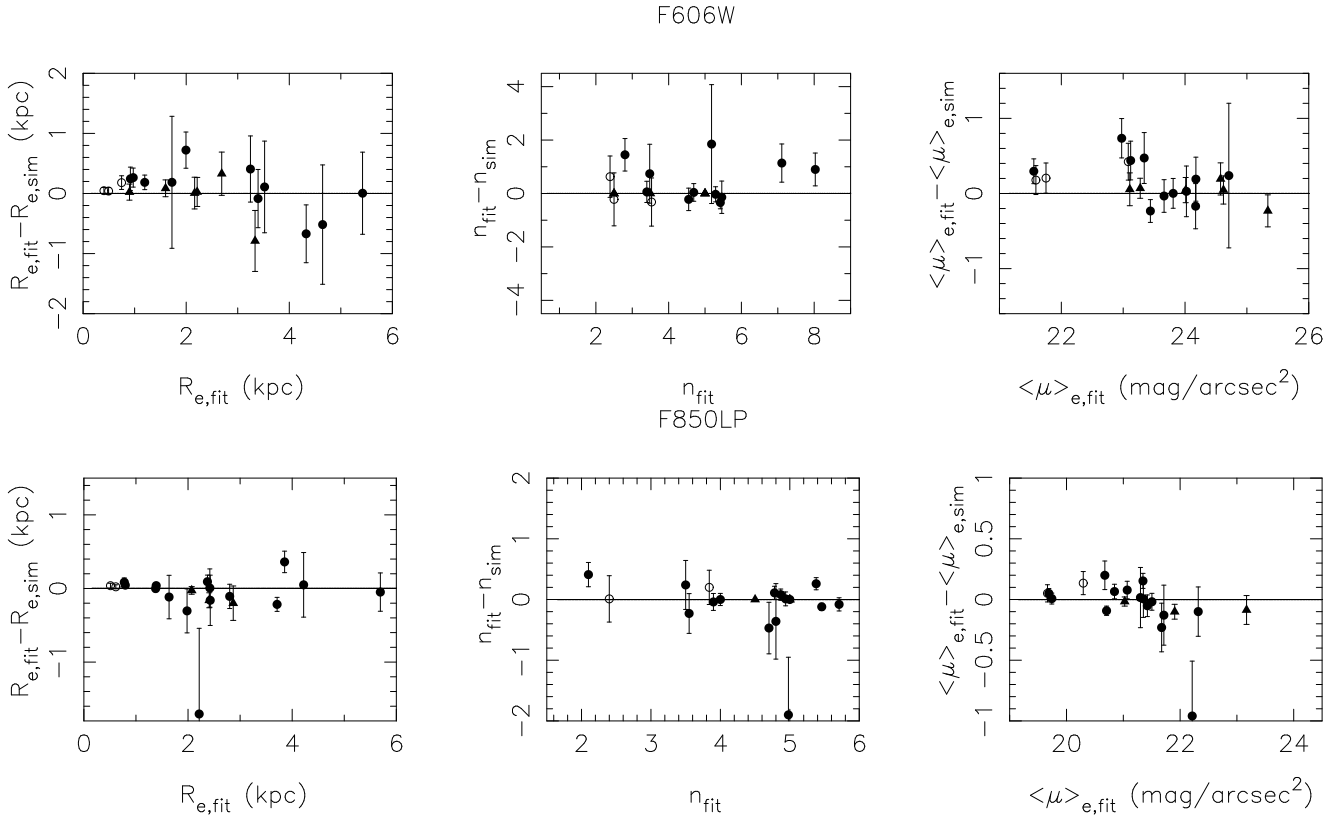

	\includegraphics[angle=-90,width=17.5cm]{figure3.ps} \\
	\includegraphics[angle=-90,width=17.5cm]{figure3_bis.ps} \\
	\caption{$Left$ $panels$ The comaprison between the estimates
          of R$_{e}$ for the real galaxies and for the simulated ones
          in the F606W band ($upper$ $panel$) and F850LP band ($lower$
          $panel$).  $Central$ $panels$ The same for Sersic index
          $n$. $Right$ $panels$ The same for the mean surface
          brightness $<\mu>_{e}$. Open points are galaxies for which
          the PSF FWHM is greater than $\sim$0.55R$_{e}$,
          while triangles represents objects with fixed Sersic index.}
	\label{simula606}
\end{figure*}

Simulations were aimed at checking for the presence of possible bias in the
measures due to instrumental effects both in the F850LP and F606W
bands. For each galaxy of our sample,with $\small{GALFIT}$, we generated 
its model with the same surface brightness parameters and magnitude and embedded
it in a real background extracted as near as possible to the real galaxy.
We recover the morphological
parameters of the simulated galaxy applying the same fitting procedure used to derive the real
profile.  The comparison between real parameters and best fitting
parameters is shown in Fig. \ref{simula606} for the two bands.
No systematic bias is detected in effective radius, Sersic index $n$, or
mean surface brightness, in either bands.

Some galaxies of our sample have an effective radius smaller than
0.1", and hence the PSF dominates the light profile beyond 0.5R$_{e}$.
Through these simulations, we assessed the reliability of
$\small{GALFIT}$ in recovering the structural parameters also for
these objects. In Fig. \ref{simula606} the open points are those
galaxies for which the ACS-PSF FWHM is greater than 0.55R$_{e}$. We chose
this limit since colour gradients are usually derived between
0.1R$_{e}$ and R$_{e}$, and hence we are interested in assessing the
accuracy of the estimate of R$_{e}$ and $n$ for all the cases in which
PSF dominates more than half of the fitted colour profile. No bias is
detected even in this case.

\begin{figure*}
	\includegraphics[angle=-90,width=17.7cm]{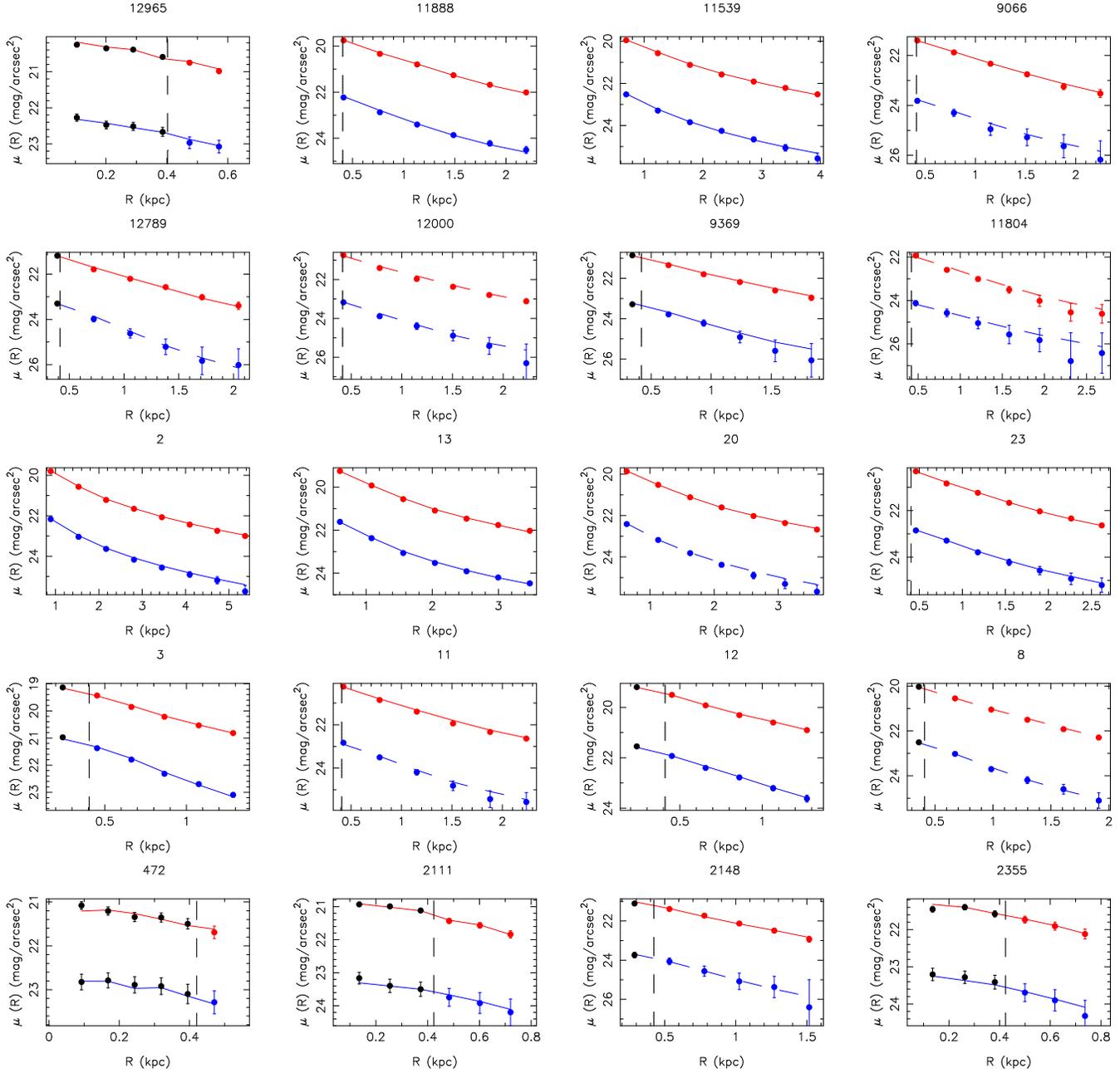} \\
	\caption{The fitted light profile convolved with the
          PSF for F606W and F850LP band (blue and red solid line in
          each panel) and the corresponding observed light profile
          (blue and red solid symbols for F606W and F850LP,
          respectively) for ETGs. The dashed vertical line set the
          radius corresponding to the PSF FWHM. Points within
          the PSF FWHM are marked by black symbols. Dashed curves
          represent galaxies with light profiles fitted with fixed
          Sersic indices.}
	\label{obsprofile1}
\end{figure*}

Finally, we compared the fitted light profile convolved with
the PSF with the observed one. This latter were computed by measuring
the surface brightness in concentric circular coronas of fixed width.
In Fig. \ref{obsprofile1} the light profiles of ETGs are
shown. Coloured lines represent the fitted
PSF-convolved light profiles for F606W and F850LP bands (blue
and red curve in each panel, respectively), and solid points trace the
corresponding profiles measured on F606W- and F850LP-band images (blue
and red points, respectively). The dashed vertical lines represent the
radius corresponding to the PSFs FWHM. Galaxies with a fixed Sersic
index are highlighted with dashed light profile curves. The errors on
the observed profiles were estimated with Sextractor \citep{bertin96}
and corrected for the correlated noise due to the drizzling following
\citet{casertano00}. This comparison can immediately help to
test the reliability of the fit.  Indeed, the agreement is generally
very good, even in the cases of compact galaxies (ID 11804, 3, 12,
472, 2111, 2148, 2355) whose light profiles are dominated by the PSF.

\subsection{Colour gradients}

\begin{figure*}
	\includegraphics[width=14.5cm,angle=-90]{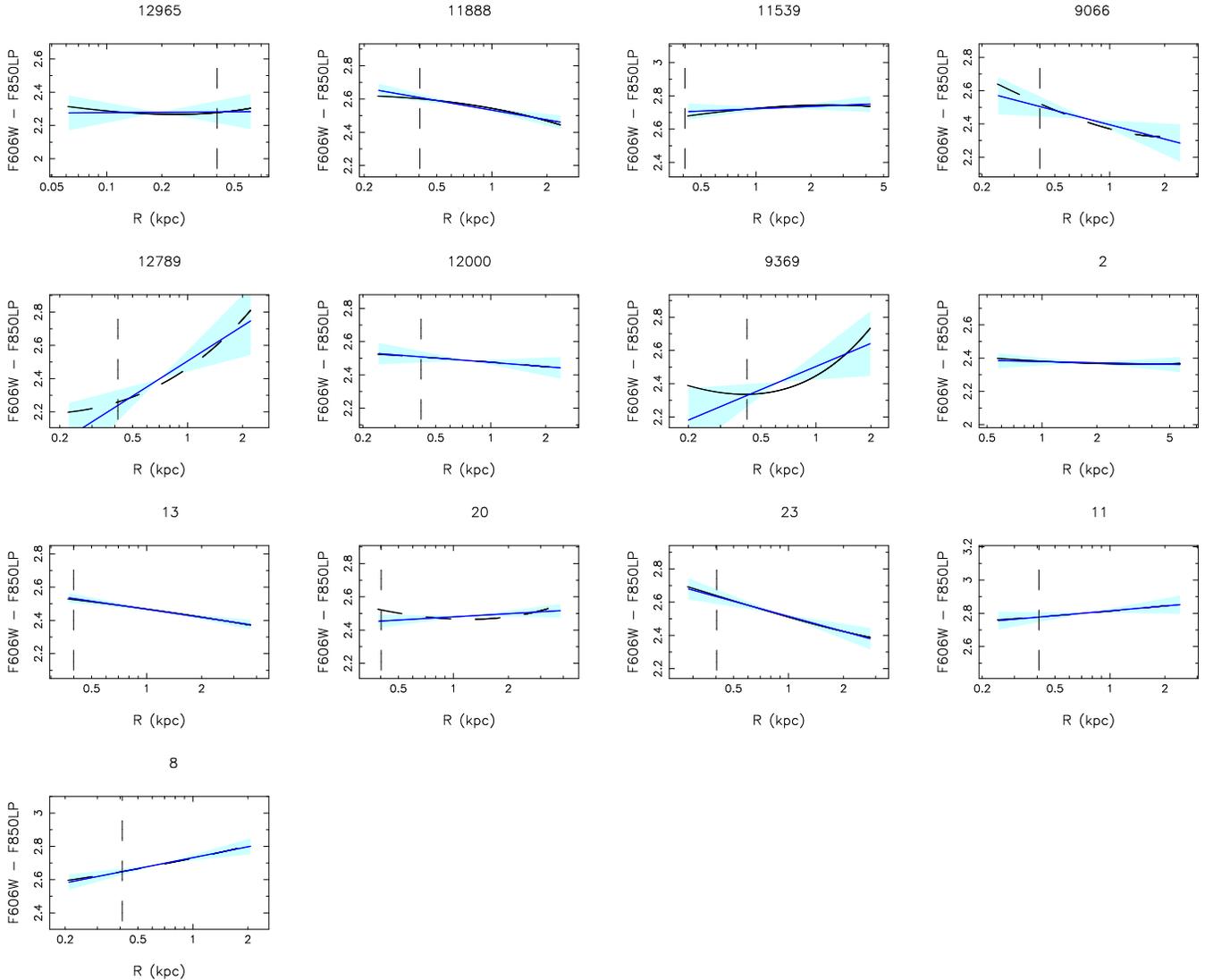} \\
	\caption{The colour gradients for ``normal" galaxies. Black
	lines represents the deconvolved colour profiles between
	0.1R$_{e}$ and R$_{e}$, and the blue lines are the best fitted
	lines to the models. The dashed vertical lines set the radius
	corresponding to the PSF FWHM. Coloured area indicate the
	1$\sigma$\, errors on colour profile slopes. Galaxies with
	light profiles fitted with a fixed Sersic index are
	represented with a dashed curve.}
	\label{cg1}
\end{figure*}

\begin{figure*}
	\includegraphics[width=7.2cm,angle=-90]{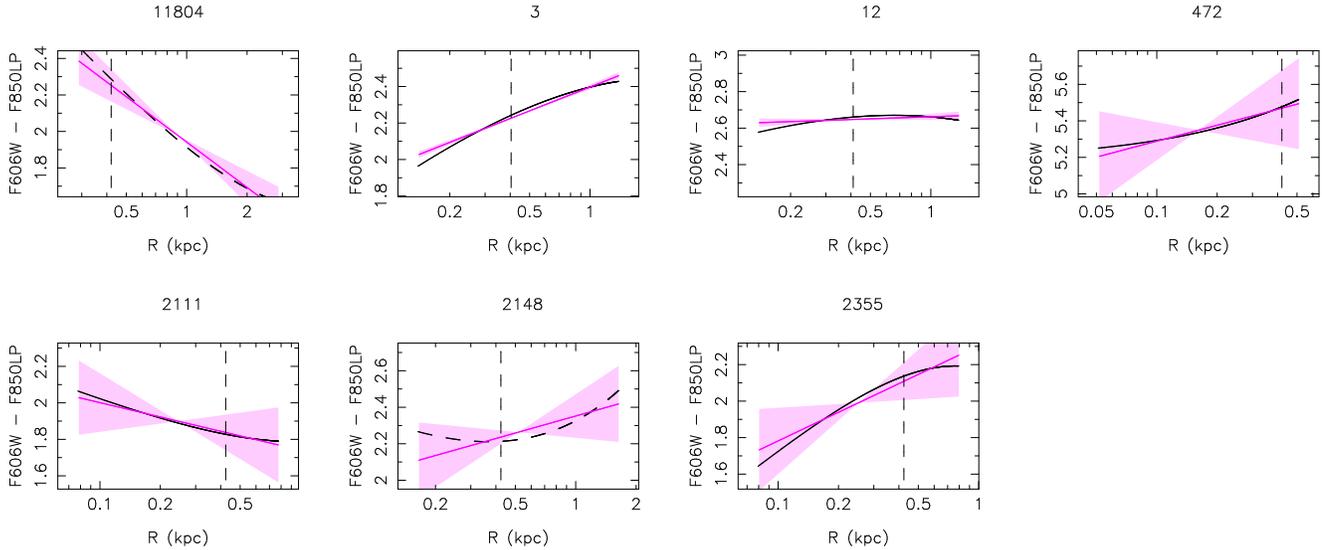} \\
	\caption{The same of Fig. \ref{cg1} but for ``compact" galaxies.}
	\label{cg2}
\end{figure*}

Following the Eq. \ref{mu}, from the structural parameters r$_{e}$, $n$,
and, M$_{tot}$, we derived the colour profiles $\mu_{UV}$-$\mu_{U}$(R)
as:
\begin{align}
\mu_{UV}-\mu_{U}(R) = \mu_{e,UV} + \frac{2.5b_{UV}}{\ln(10)}[(R/R_{e,UV})^{1/n_{UV}}-1] + \\ 
- \mu_{e,U} - \frac{2.5b_{U}}{\ln(10)}[(R/R_{e,U})^{1/n_{U}}-1]
\end{align}
and estimated their logarithmic slope with an orthogonal least-squares
fit.  To be consistent with the previous studies on colour gradients
(see, e.g, Peletier et al. 1990) of local ETGs we fitted the profiles
between 0.1R$_{e}$ and R$_{e}$. In Fig \ref{cg1} and \ref{cg2} the
results are reported for ``normal" and ``compact" galaxies,
respectively.
Black curves represent the colour profiles between 0.1R$_{e}$ and
R$_{e}$ and coloured lines are the fitted slopes. The dashed vertical lines
indicate the radius of the PSF at FWHM. Black dashed curves represent galaxies with
light profiles fitted with a Sersic index fixed for at least one band.
The coloured area indicate the 1$\sigma$\,errors on colour gradient.
For each galaxy, in Tab.\,\ref{sp} the value of the gradient and its
corresponding error are reported. The latter one is not the formal
error provided by $\small{GALFIT}$ for structural parameters (see
Table \ref{sp}) but takes into account also the effect of S/N on the
estimated parameters.
 
\begin{figure*}
	\includegraphics[width=190pt,angle=-90]{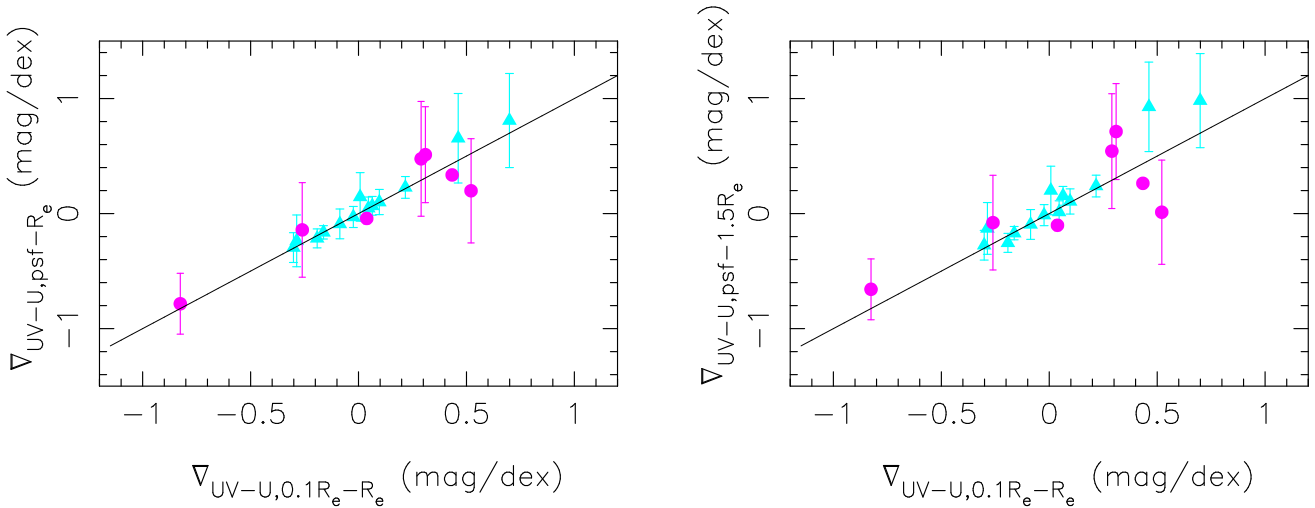} \\
	\caption{$Left$ $Panel$: Comparison between ``classical"
	colour gradient estimated between 0.1R$_{e}$ and R$_{e}$
	($\nabla_{UV-U,0.1R_{e}-R_{e}}$) and those estimated between
	the radius corresponding to the ACS-PSF FWHM (0.05") and
	R$_{e}$ ($\nabla_{UV-U,psf-R_{e}}$). $Right$ $Panel$:
	Comparison between ``classical" colour gradient estimated
	between 0.1R$_{e}$ and R$_{e}$ and those estimated between the
	radius corresponding to the ACS-PSF FWHM(0.05") and
	1.5R$_{e}$. Solid triangles are normal galaxies, while solid
	circles are compact ones.}
	\label{cgpsf}
\end{figure*}

To this aim, for each galaxy, we generated the corresponding simulated
galaxy both in F606W and F850LP band and embedded each of them in 50
different backgrounds extracted from their own image. For each galaxy,
considering all the possible combinations between these two set of
simulations, we measured 2500 simulated colour gradients and set the
standard deviation of this sample as final error on the real colour
gradient.  From the Fig. \ref{cg1} and \ref{cg2}, it immediately turns
out that despite the extremely narrow wavelength baseline covered
(1000 \AA \, in the rest-frame) which samples approximately the same
stellar population, we detected colour variations in high-$z$ ETGs.  The
result is not dependent by the fitted region. Indeed, we re-estimated
the colour gradients between the radius of the PSF FWHM and R$_{e}$
and between the radius of the PSF FWHM and 1.5R$_{e}$ and compared
them with the ``classical" ones between 0.1R$_{e}$ and R$_{e}$
(Fig. \ref{cgpsf}).
The three estimates are consistent each other and even fitting the
gradients outside the radius corresponding to the PSF FWHM, ETGs still
show both positive and negative variation in the colour along their
radius.

As pointed out before, colour gradients for five galaxies of our
sample were derived with the Sersic index fixed in the fitting of the
F606W band, while for three galaxies the structural parameter was
locked in the fitting of both images. To test how much a misleading
estimate of the $n$ parameter can affect the gradient, for each ETG
with Sersic index locked in at least one band, we simulated 14
galaxies with structural parameters equal to the real one and Sersic
index varying in the range [1.75$\div$8.25] with a step of
0.5. We added real background noise to these models and we fit
the light profiles of the simulated galaxies holding $n$ fixed to the
original value (Table 1) while allowing the other parameters to
vary. From the analysis of the corresponding colour gradients, it
turns out that for galaxies with Sersic index locked in a single band,
the colour gradients of simulated galaxies with varying $n$ are fully
compatible with the value we derived having set the Sersic index to
the value we present. Concurrently, simulations suggest that the
gradient estimate is not so stable to the simultaneous variation of
the $n$ in both bands. Thus, our estimates of colour gradients result
to be fully reliable for 17 out of 20 galaxies.

\section{Results}

From Fig.\,\ref{cg1} and Fig.\,\ref{cg2} it clearly emerges that,
despite the narrow wavelength baseline covered, high-$z$ ETGs show
significant colour gradients. Most intriguingly, differently from what
is widely found in the local Universe, high-$z$ ETGs do not show a
systematic reddening toward the center, but exhibit a uniform
distribution of gradients from positive to negative values as can also
be seen in Fig \ref{cgpsf}. Indeed, while 10 out of 20 galaxies have
colour gradients consistent with a flat distribution of blue stars
within high-$z$ ETGs (null colour gradient, galaxy ID: 12965, 11539,
12000, 2, 20, 11, 12, 2148, 2111, 472), five galaxies show effective
negative colour gradients (galaxy ID: 11888, 9066, 11804, 13, 23) and
the remaining five have sharp positive gradients (galaxy ID: 12789,
9369, 3, 8, 2355). This spread from positive to negative values
reflects into a mean colour gradient for the whole sample of $\sim$
$<$$\nabla$$_{UV-U}$$>$\,=\, 0.10$\pm$0.30 mag\,dex$^{-1}$. Although
we present the widest sample of high-$z$ colour gradients available at
the moment, we remind the reader that it is not complete in magnitude
and/or mass and this limitation can affect the value of the mean
colour gradient $<$$\nabla$$_{UV-U}$$>$ for the whole population of
high-$z$ ETGs.  At the same time, while the incompleteness of our
sample can affect the mean value, it does not influence the other
important result we obtain: the spatial distribution of stellar
content is not univocal in the high-$z$ ETGs population. Indeed, our
results show that high-$z$ ETGs are a composite population formed by
galaxies with the bluest population placed either in the center or in
the periphery as well as distributed homogeneously throughout the
whole galaxy.

\begin{figure}
	\includegraphics[width=160pt,angle=-90]{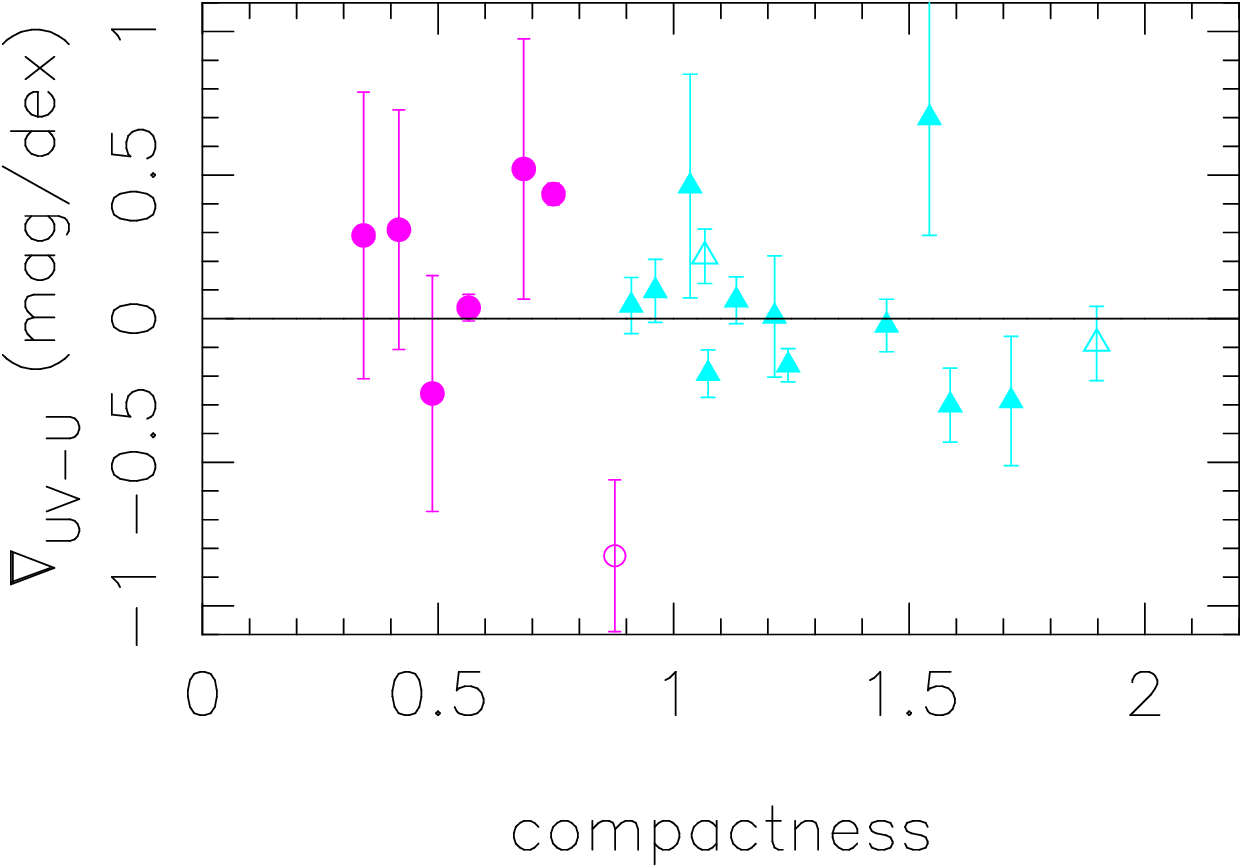} \\
	\caption{Colour gradient for our sample of galaxies as a
	function of compactness defined as the ratio
	R$_{e}$/(R$_{e,z=0}$-1\,$\sigma$) where R$_{e}$ is the
	effective radius of the galaxy and R$_{e,z=0}$ is the
	effective radius that a galaxy of equal mass would have at
	$z$\,=\,0 as derived by the local size-mass (SM) relation. We
	define compact those galaxies with R$_e$ more than one sigma
	smaller than those predicted by local SM relation for that
	mass, which corresponds to values of the compactness
	R$_{e}$/(R$_{e,z=0}$-1\,$\sigma$) $<$ 0.9. Circles are
	compact galaxies, while triangles are the normal ones. Open
	points are galaxies with Sersic index fixed in the fit on both
	bands.}
	\label{gradcomp}
\end{figure}

To assess whether any difference exists in the stellar distribution of
normal and compact high-$z$ ETGs, in Fig. \ref{gradcomp} we report
colour gradient as a function of the galaxy degree of compactness
(see Sect. 2). Circles represent compact galaxies, while
triangles are normal ones. The open points represent galaxies for
which Sersic indices are locked in both bands. Since simulations leave
a room of space to uncertainties into the estimates of the colour
gradients in these systems, we prefer not to take into account these
points.  We observe that the present galaxy sample seems to show a
mild correlation between the degree of compactness and colour
gradient.  Indeed, compact galaxies seem to preferentially show a
bluer core than the outer regions, while moving towards normal
galaxies stellar populations become redder in the center as observed
in the local Universe.  In fact, only one compact galaxy out of 6
(15$\%$) has a negative colour gradient (but compatible within the
error with a positive value) against the 45$\%$ in the case of normal
galaxies.  This difference is detectable in the mean colour gradient
value of the two samples.  Indeed, while the mean colour gradient of
compact galaxies is $<$$\nabla$$_{UV-U}$$>_{comp}$\,=\, 0.22$\pm$0.28,
normal galaxies have a mean value of
$<$$\nabla$$_{UV-U}$$>_{norm}$\,=\, 0.04$\pm$0.30. Even if the two
mean estimates seem to point towards a different nature of the two
samples, it is to note that they are consistent within the errors and
that the KS probability for the two samples of being extracted from
different populations is not significant ($\sim$26$\%$).  At the same
time, the Spearman's rank test with a coefficient $\rho$\,=\,-0.34
shows that the correlation between the two quantities has a
probability of $\sim$ 85$\%$ to not be observed by chance, but even in
this case, the poor statistics of our sample prevent us to reach a
firm conclusion. In fact, the trend slightly hinted in
Fig. \ref{gradcomp} can be even due to the incompleteness of our
sample. Unfortunately, we are not able to discriminate how and in
which direction (if any) the selection criteria of our sample can
arrange the distribution of the points in Fig. \ref{gradcomp}, and
hence the effect they could have on the supposed connection.  Thus,
although the quality of our sample allow us to investigate the
internal colour distribution of both compact and
normal high-$z$ ETGs, we cannot be conclusive about the suggested
physical connection between the degree of compactness and radial
colour variations.

\section{Conclusions}

We analysed (UV-U)$_{rest frame}$ colour gradients in a sample of 20
ETGs at 0.9$<z_{spec}<$1.92 comprising both normal and compact galaxies.
Despite the short wavelength baseline covered, we detect effective
radial colour variations in $\sim$ 50$\%$ of the galaxies of our sample,
while the remaining show a gradient comparable with zero. In
particular, we detect significant radial colour variations in 10 out of the 20
galaxies of our sample: five galaxies exhibit a reddening towards
the internal regions (negative colour gradient), as generally observed
in the local Universe, and five galaxies show stellar populations
bluer in the central regions than in the outskirts, which reflects in
positive colour gradients. This result supports the previous findings on
colour gradients at intermediate and high redshifts
\citep[e.g.][]{menanteau01,moth02,ferreras09} which already revealed the presence
of blue core ETGs which are not common in the local Universe.

Taking advantage of our composite sample including both compact and
normal ETGs, we investigate possible systematics in the stellar
content of these physically different systems. Actually, the
distribution of our data in the plane defined by colour gradients and
compactness is suggestive of a mild correlation whereby the colour
gradient decreases from positive to negative values moving from
compact towards normal galaxies.  This trend reflects in the mean
values of colour gradients of compact and normal galaxies,
0.22$\pm$0.28 and 0.04$\pm$0.30 respectively, even if the difference
is not statistically significant to point toward two strictly
different mass assembly histories.  Concurrently, the Spearman's test
assess that the correlation between ETGs compactness and colour
gradient could be present with a probability of $\sim$ 85$\%$
suggesting a possible physical connection between the two quantities.
However, we cannot reach a firm conclusion since the poor statistics
and the incompleteness could concur to originate a fake
correlation. Moreover, it is to note that this trend could be affected
by the short wavelength baseline covered which do not sample the
emission of the oldest stellar populations.

Although the limits of our sample do not allow us to strictly assess
or deny this suggested trend, theoretical simulations of the
hierarchical assembly of the stellar matter seem to support the
presence of blue cores in compact galaxies \citep{hopkins08}. Indeed,
at higher redshifts, the progenitor galaxies tend to have a greater
amount of gas not yet converted in stars. In a merger event involving
a gas rich galaxy, the torsion forces convey this reservoir toward the
center giving origin to an intense and short burst of star
formation. Concurrently, the pre-existing stars tend to arrange
themselves in the external regions leading to the formation of an ETG
with a younger (and hence bluer) stellar population in the center than
in the outskirts and, consequently, with a positive colour gradient. On
the contrary, in later epochs, the progenitor galaxies, having already
consumed their gas both in internal processes and in previous mergers,
cannot fuel the central starburst and, hence, their dry merger will
not result in a blue core galaxy. Additionally, the enhancement with
redshift of the gas fraction at the disposal of progenitor galaxies could
explain not only the increasing presence of blue core galaxies at
higher redshift, but also their compact dimensions. In fact, the blue
core galaxies, due to the central starburst they underwent in the
early strong dissipative merger, will have surface brightness profiles
more centrally peaked, hence with smaller R$_{e}$, than ETGs formed in
non-dissipative mergers. As a consequence, ETGs formed in early mergers
will tend to have both bluer cores and to be more compact.

Supporting this scenario, recently it has been found that high-$z$ ETGs
formed at earlier epochs are preferentially compact galaxies \citep{saracco10b}. 
Even if these evidences seem to concur to this picture,
many aspect are still missing, e.g. which internal and/or external
processes would be responsible of the enhanced central star formation,
and hence of the positive colour gradients, observed also in normal
galaxies.

To gain more insight in the complex scenario of early-type galaxy
formation and evolution it would be fruitful enlarge the wavelength
baseline covered, in order to sample also the emission dominated by the oldest stellar
population (Gargiulo et al. in preparation). 
This will allow us to better characterize
the stellar content of high-$z$ ETGs as whole and, in particular, to
deeply investigate the presence of possible systematics between the
stellar populations of normal and compact ETGs.

\section*{Acknowledgments}

I would like to thank Chris Haines and Francesco La Barbera for their
helpful comments and support.  This work is based on observations made
with the NASA/ESA Hubble Space Telescope, obtained from the data
archive at the Space Telescope Science Institute which is operated by
the Association of Universities for Research in Astronomy. This work
has received financial support from ASI (contract I/016/07/0 and
I/009/10/0).

\nocite{}
\bibliographystyle{mn2e}
\bibliography{gargiulo_vreferee2}

\begin{thebibliography}{}

\bibitem[\protect\citeauthoryear{{Bertin} \& {Arnouts}}{{Bertin} \&
  {Arnouts}}{1996}]{bertin96}
{Bertin} E.,  {Arnouts} S.,  1996, A$\&$AS, 117, 393

\bibitem[\protect\citeauthoryear{{Boylan-Kolchin}, {Ma} \&
  {Quataert}}{{Boylan-Kolchin} et~al.}{2006}]{boylan06}
{Boylan-Kolchin} M.,  {Ma} C.,    {Quataert} E.,  2006, MNRAS, 369, 1081

\bibitem[\protect\citeauthoryear{{Cappellari}, {di Serego Alighieri},
  {Cimatti}, {Daddi}, {Renzini}, {Kurk}, {Cassata}, {Dickinson},
  {Franceschini}, {Mignoli}, {Pozzetti}, {Rodighiero}, {Rosati} \&
  {Zamorani}}{{Cappellari} et~al.}{2009}]{cappellari09}
{Cappellari} M.,  {di Serego Alighieri} S.,  {Cimatti} A.,  {Daddi} E.,
  {Renzini} A.,  {Kurk} J.~D.,  {Cassata} P.,  {Dickinson} M.,  {Franceschini}
  A.,  {Mignoli} M.,  {Pozzetti} L.,  {Rodighiero} G.,  {Rosati} P.,
  {Zamorani} G.,  2009, ApJL, 704, L34

\bibitem[\protect\citeauthoryear{{Casertano}, {de Mello}, {Dickinson},
  {Ferguson}, {Fruchter}, {Gonzalez-Lopezlira}, {Heyer}, {Hook}, {Levay},
  {Lucas}, {Mack}, {Makidon}, {Mutchler}, {Smith}, {Stiavelli}, {Wiggs} \&
  {Williams}}{{Casertano} et~al.}{2000}]{casertano00}
{Casertano} S.,  {de Mello} D.,  {Dickinson} M.,  {Ferguson} H.~C.,  {Fruchter}
  A.~S.,  {Gonzalez-Lopezlira} R.~A.,  {Heyer} I.,  {Hook} R.~N.,  {Levay} Z.,
  {Lucas} R.~A.,  {Mack} J.,  {Makidon} R.~B.,  {Mutchler} M.,  {Smith} T.~E.,
  {Stiavelli} M.,  {Wiggs} M.~S.,    {Williams} R.~E.,  2000, AJ, 120, 2747

\bibitem[\protect\citeauthoryear{{Cenarro} \& {Trujillo}}{{Cenarro} \&
  {Trujillo}}{2009}]{cenarro09}
{Cenarro} A.~J.,  {Trujillo} I.,  2009, ApJL, 696, L43

\bibitem[\protect\citeauthoryear{{Cimatti}, {Cassata}, {Pozzetti}, {Kurk},
  {Mignoli}, {Renzini}, {Daddi}, {Bolzonella}, {Brusa}, {Rodighiero},
  {Dickinson}, {Franceschini}, {Zamorani}, {Berta}, {Rosati} \&
  {Halliday}}{{Cimatti} et~al.}{2008}]{cimatti08}
{Cimatti} A.,  {Cassata} P.,  {Pozzetti} L.,  {Kurk} J.,  {Mignoli} M.,
  {Renzini} A.,  {Daddi} E.,  {Bolzonella} M.,  {Brusa} M.,  {Rodighiero} G.,
  {Dickinson} M.,  {Franceschini} A.,  {Zamorani} G.,  {Berta} S.,  {Rosati}
  P.,    {Halliday} C.,  2008, A$\&$A, 482, 21

\bibitem[\protect\citeauthoryear{{Damjanov}, {McCarthy}, {Abraham},
  {Glazebrook}, {Yan}, {Mentuch}, {Le Borgne}, {Savaglio}, {Crampton},
  {Murowinski}, {Juneau}, {Carlberg}, {J{\o}rgensen}, {Roth}, {Chen} \&
  {Marzke}}{{Damjanov} et~al.}{2009}]{damjanov09}
{Damjanov} I.,  {McCarthy} P.~J.,  {Abraham} R.~G.,  {Glazebrook} K.,  {Yan}
  H.,  {Mentuch} E.,  {Le Borgne} D.,  {Savaglio} S.,  {Crampton} D.,
  {Murowinski} R.,  {Juneau} S.,  {Carlberg} R.~G.,  {J{\o}rgensen} I.,  {Roth}
  K.,  {Chen} H.,    {Marzke} R.~O.,  2009, ApJ, 695, 101

\bibitem[\protect\citeauthoryear{{di Serego Alighieri}, {Vernet}, {Cimatti},
  {Lanzoni}, {Cassata}, {Ciotti}, {Daddi}, {Mignoli}, {Pignatelli}, {Pozzetti},
  {Renzini}, {Rettura} \& {Zamorani}}{{di Serego Alighieri}
  et~al.}{2005}]{diserego05}
{di Serego Alighieri} S.,  {Vernet} J.,  {Cimatti} A.,  {Lanzoni} B.,
  {Cassata} P.,  {Ciotti} L.,  {Daddi} E.,  {Mignoli} M.,  {Pignatelli} E.,
  {Pozzetti} L.,  {Renzini} A.,  {Rettura} A.,    {Zamorani} G.,  2005, A$\&$A,
  442, 125

\bibitem[\protect\citeauthoryear{{Fan}, {Lapi}, {De Zotti} \& {Danese}}{{Fan}
  et~al.}{2008}]{fan08}
{Fan} L.,  {Lapi} A.,  {De Zotti} G.,    {Danese} L.,  2008, ApJ, 689, L101

\bibitem[\protect\citeauthoryear{{Ferreras}, {Lisker}, {Pasquali} \&
  {Kaviraj}}{{Ferreras} et~al.}{2009}]{ferreras09}
{Ferreras} I.,  {Lisker} T.,  {Pasquali} A.,    {Kaviraj} S.,  2009, MNRAS,
  395, 554

\bibitem[\protect\citeauthoryear{{Giavalisco}, {Ferguson}, {Koekemoer},
  {Dickinson}, {Alexander}, {Bauer}, {Bergeron}, {Biagetti}, {Brandt} \&
  {Casertano}}{{Giavalisco} et~al.}{2004}]{giavalisco04}
{Giavalisco} M.,  {Ferguson} H.~C.,  {Koekemoer} A.~M.,  {Dickinson} M.,
  {Alexander} D.~M.,  {Bauer} F.~E.,  {Bergeron} J.,  {Biagetti} C.,  {Brandt}
  W.~N.,    {Casertano} S.,  2004, ApJ, 600, L93

\bibitem[\protect\citeauthoryear{{H{\"a}ussler}, {McIntosh}, {Barden}, {Bell},
  {Rix}, {Borch}, {Beckwith}, {Caldwell}, {Heymans}, {Jahnke}, {Jogee},
  {Koposov}, {Meisenheimer}, {S{\'a}nchez}, {Somerville}, {Wisotzki} \&
  {Wolf}}{{H{\"a}ussler} et~al.}{2007}]{haussler07}
{H{\"a}ussler} B.,  {McIntosh} D.~H.,  {Barden} M.,  {Bell} E.~F.,  {Rix} H.,
  {Borch} A.,  {Beckwith} S.~V.~W.,  {Caldwell} J.~A.~R.,  {Heymans} C.,
  {Jahnke} K.,  {Jogee} S.,  {Koposov} S.~E.,  {Meisenheimer} K.,
  {S{\'a}nchez} S.~F.,  {Somerville} R.~S.,  {Wisotzki} L.,    {Wolf} C.,
  2007, ApJS, 172, 615

\bibitem[\protect\citeauthoryear{{Hopkins}, {Bundy}, {Hernquist}, {Wuyts} \&
  {Cox}}{{Hopkins} et~al.}{2010}]{hopkins10}
{Hopkins} P.~F.,  {Bundy} K.,  {Hernquist} L.,  {Wuyts} S.,    {Cox} T.~J.,
  2010, MNRAS, 401, 1099

\bibitem[\protect\citeauthoryear{{Hopkins}, {Cox} \& {Hernquist}}{{Hopkins}
  et~al.}{2008}]{hopkins08}
{Hopkins} P.~F.,  {Cox} T.~J.,    {Hernquist} L.,  2008, ApJ, 689, 17

\bibitem[\protect\citeauthoryear{{La Barbera} \& {de Carvalho}}{{La Barbera} \&
  {de Carvalho}}{2009}]{labarbera09}
{La Barbera} F.,  {de Carvalho} R.~R.,  2009, ApJ, 699, L76

\bibitem[\protect\citeauthoryear{{Mancini}, {Daddi}, {Renzini}, {Salmi},
  {McCracken}, {Cimatti}, {Onodera}, {Salvato}, {Koekemoer}, {Aussel}, {Le
  Floc'h}, {Willott} \& {Capak}}{{Mancini} et~al.}{2010}]{mancini10}
{Mancini} C.,  {Daddi} E.,  {Renzini} A.,  {Salmi} F.,  {McCracken} H.~J.,
  {Cimatti} A.,  {Onodera} M.,  {Salvato} M.,  {Koekemoer} A.~M.,  {Aussel} H.,
   {Le Floc'h} E.,  {Willott} C.,    {Capak} P.,  2010, MNRAS, 401, 933

\bibitem[\protect\citeauthoryear{{McGrath}, {Stockton}, {Canalizo}, {Iye} \&
  {Maihara}}{{McGrath} et~al.}{2008}]{mcgrath08}
{McGrath} E.~J.,  {Stockton} A.,  {Canalizo} G.,  {Iye} M.,    {Maihara} T.,
  2008, ApJ, 682, 303

\bibitem[\protect\citeauthoryear{{Menanteau}, {Abraham} \& {Ellis}}{{Menanteau}
  et~al.}{2001}]{menanteau01}
{Menanteau} F.,  {Abraham} R.~G.,    {Ellis} R.~S.,  2001, MNRAS, 322, 1

\bibitem[\protect\citeauthoryear{{Menanteau}, {Martel}, {Tozzi}, {Frye},
  {Ford}, {Infante}, {Ben{\'{\i}}tez}, {Galaz}, {Coe}, {Illingworth}, {Hartig}
  \& {Clampin}}{{Menanteau} et~al.}{2005}]{menanteau05}
{Menanteau} F.,  {Martel} A.~R.,  {Tozzi} P.,  {Frye} B.,  {Ford} H.~C.,
  {Infante} L.,  {Ben{\'{\i}}tez} N.,  {Galaz} G.,  {Coe} D.,  {Illingworth}
  G.~D.,  {Hartig} G.~F.,    {Clampin} M.,  2005, ApJ, 620, 697

\bibitem[\protect\citeauthoryear{{Moth} \& {Elston}}{{Moth} \&
  {Elston}}{2002}]{moth02}
{Moth} P.,  {Elston} R.~J.,  2002, AJ, 124, 1886

\bibitem[\protect\citeauthoryear{{Naab}, {Johansson} \& {Ostriker}}{{Naab}
  et~al.}{2009}]{naab09}
{Naab} T.,  {Johansson} P.~H.,    {Ostriker} J.~P.,  2009, ApJ, 699, L178

\bibitem[\protect\citeauthoryear{{Nipoti}, {Treu}, {Auger} \&
  {Bolton}}{{Nipoti} et~al.}{2009}]{nipoti09}
{Nipoti} C.,  {Treu} T.,  {Auger} M.~W.,    {Bolton} A.~S.,  2009, ApJ, 706,
  L86

\bibitem[\protect\citeauthoryear{{Onodera}, {Daddi}, {Gobat}, {Cappellari},
  {Arimoto}, {Renzini}, {Yamada}, {McCracken}, {Mancini} \& {Capak}}{{Onodera}
  et~al.}{2010}]{onodera10}
{Onodera} M.,  {Daddi} E.,  {Gobat} R.,  {Cappellari} M.,  {Arimoto} N.,
  {Renzini} A.,  {Yamada} Y.,  {McCracken} H.~J.,  {Mancini} C.,    {Capak} P.,
   2010, ApJL, 715, L6

\bibitem[\protect\citeauthoryear{{Peletier}, {Valentijn} \&
  {Jameson}}{{Peletier} et~al.}{1990}]{peletier90}
{Peletier} R.~F.,  {Valentijn} E.~A.,    {Jameson} R.~F.,  1990, A$\&$A, 233,
  62

\bibitem[\protect\citeauthoryear{{Peng}, {Ho}, {Impey} \& {Rix}}{{Peng}
  et~al.}{2002}]{peng02}
{Peng} C.~Y.,  {Ho} L.~C.,  {Impey} C.~D.,    {Rix} H.,  2002, AJ, 124, 266

\bibitem[\protect\citeauthoryear{{Rettura}, {Rosati}, {Strazzullo}, {Dickinson}
 {Fosbury}, {Rocca-Volmerange}, {Cimatti}, {di Serego Alighieri}, {Kuntschner},
{Lanzoni}, {Nonino}, {Popesso}, {Stern}, {Eisenhardt}, {Lidman}, \& {Stanford}}
{{Rettura} et~al.}{2006}]{rettura06}
{Rettura} A., {Rosati} P., {Strazzullo} V., {Dickinson} M., {Fosbury} R.~A.~E.,
{Rocca-Volmerange} B., {Cimatti} A., {di Serego Alighieri} S., {Kuntschner} H.,
{Lanzoni} B., {Nonino} M., {Popesso} P., {Stern} D., {Eisenhardt} P.~R., 
{Lidman} C., {Stanford}, S.~A., 2006, A$\&$A, 395, 554

\bibitem[\protect\citeauthoryear{{Saglia}, {Maraston}, {Greggio}, {Bender} \&
  {Ziegler}}{{Saglia} et~al.}{2000}]{saglia00}
{Saglia} R.~P.,  {Maraston} C.,  {Greggio} L.,  {Bender} R.,    {Ziegler} B.,
  2000, A$\&$A, 360, 911

\bibitem[\protect\citeauthoryear{{Saracco}, {Longhetti} \& {Andreon}}{{Saracco}
  et~al.}{2009}]{saracco09}
{Saracco} P.,  {Longhetti} M.,    {Andreon} S.,  2009, MNRAS, 392, 718

\bibitem[\protect\citeauthoryear{{Saracco}, {Longhetti} \&
  {Gargiulo}}{{Saracco} et~al.}{2010a}]{saracco10}
{Saracco} P.,  {Longhetti} M.,    {Gargiulo} A.,  2010a, MNRAS, pp L115+

\bibitem[\protect\citeauthoryear{{Saracco}, {Longhetti} \&
  {Gargiulo}}{{Saracco} et~al.}{2010b}]{saracco10b}
{Saracco} P.,  {Longhetti} M.,    {Gargiulo} A.,  2010b, MNRAS, submitted

\bibitem[\protect\citeauthoryear{{Sersic}}{{Sersic}}{1968}]{sersic68}
{Sersic} J.~L.,  1968, {Atlas de galaxias australes}

\bibitem[\protect\citeauthoryear{{Shen}, {Mo}, {White}, {Blanton}, {Kauffmann},
  {Voges}, {Brinkmann} \& {Csabai}}{{Shen} et~al.}{2003}]{shen03}
{Shen} S.,  {Mo} H.~J.,  {White} S.~D.~M.,  {Blanton} M.~R.,  {Kauffmann} G.,
  {Voges} W.,  {Brinkmann} J.,    {Csabai} I.,  2003, MNRAS, 343, 978

\bibitem[\protect\citeauthoryear{{Stockton}, {Shih} \& {Larson}}{{Stockton}
  et~al.}{2010}]{stockton10}
{Stockton} A.,  {Shih} H.,    {Larson} K.,  2010, ApJL, 709, L58

\bibitem[\protect\citeauthoryear{{Trujillo}, {Cenarro}, {de
  Lorenzo-C{\'a}ceres}, {Vazdekis}, {de la Rosa} \& {Cava}}{{Trujillo}
  et~al.}{2009}]{trujillo09}
{Trujillo} I.,  {Cenarro} A.~J.,  {de Lorenzo-C{\'a}ceres} A.,  {Vazdekis} A.,
  {de la Rosa} I.~G.,    {Cava} A.,  2009, ApJ, 692, L118

\bibitem[\protect\citeauthoryear{{Valentinuzzi}, {Fritz}, {Poggianti}, {Cava},
  {Bettoni}, {Fasano}, {D'Onofrio}, {Couch}, {Dressler}, {Moles}, {Moretti},
  {Omizzolo}, {Kj{\ae}rgaard}, {Vanzella} \& {Varela}}{{Valentinuzzi}
  et~al.}{2010a}]{valentinuzzi10}
{Valentinuzzi} T.,  {Fritz} J.,  {Poggianti} B.~M.,  {Cava} A.,  {Bettoni} D.,
  {Fasano} G.,  {D'Onofrio} M.,  {Couch} W.~J.,  {Dressler} A.,  {Moles} M.,
  {Moretti} A.,  {Omizzolo} A.,  {Kj{\ae}rgaard} P.,  {Vanzella} E.,
  {Varela} J.,  2010a, ApJ, 712, 226

\bibitem[\protect\citeauthoryear{{Valentinuzzi}, {Poggianti}, {Saglia},
  {Aragon-Salamanca}, {Simard}, {Sanchez-Blazquez}, {D'Onofrio}, {Cava},
  {Couch}, {Fritz}, {Moretti} \& {Vulcani}}{{Valentinuzzi}
  et~al.}{2010b}]{valentinuzzi10b}
{Valentinuzzi} T.,  {Poggianti} B.~M.,  {Saglia} R.~P.,  {Aragon-Salamanca} A.,
   {Simard} L.,  {Sanchez-Blazquez} P.,  {D'Onofrio} M.,  {Cava} A.,  {Couch}
  W.~J.,  {Fritz} J.,  {Moretti} A.,    {Vulcani} B.,  2010b, ArXiv e-prints

\bibitem[\protect\citeauthoryear{{van der Wel}, {Franx}, {van Dokkum}, {Rix},
  {Illingworth} \& {Rosati}}{{van der Wel} et~al.}{2005}]{vanderwel05}
{van der Wel} A.,  {Franx} M.,  {van Dokkum} P.~G.,  {Rix} H.,  {Illingworth}
  G.~D.,    {Rosati} P.,  2005, ApJ, 631, 145

\bibitem[\protect\citeauthoryear{{van Dokkum}, {Kriek} \& {Franx}}{{van Dokkum}
  et~al.}{2009}]{vandokkum09}
{van Dokkum} P.~G.,  {Kriek} M.,    {Franx} M.,  2009, Nature, 460, 717

\bibitem[\protect\citeauthoryear{{Vanzella}, {Cristiani}, {Dickinson},
  {Giavalisco}, {Kuntschner}, {Haase}, {Nonino}, {Rosati}, {Cesarsky},
  {Ferguson} \& {GOODS Team}}{{Vanzella} et~al.}{2008}]{vanzella08}
{Vanzella} E.,  {Cristiani} S.,  {Dickinson} M.,  {Giavalisco} M.,
  {Kuntschner} H.,  {Haase} J.,  {Nonino} M.,  {Rosati} P.,  {Cesarsky} C.,
  {Ferguson} H.~C.,    {GOODS Team} 2008, A$\&$A, 478, 83

\end{thebibliography}

\end{document}